\documentclass[a4paper,aps,prl,reprint,superscriptaddress]{revtex4-1}

\usepackage{amsmath} 

\usepackage{graphicx}
\usepackage[colorlinks,
	linkcolor=red,
	citecolor=blue,
	urlcolor=red]{hyperref}
\usepackage[all]{hypcap}

\usepackage{cleveref}

\usepackage{layouts}

\usepackage{times}

\usepackage[english]{babel}

\usepackage[xcolor={usenames,dvipsnames}]{changes}
\definechangesauthor[color=purple]{NRB}
\definechangesauthor[color=blue]{TJK}
\definechangesauthor[color=Green]{LDT}
\definechangesauthor[color=ProcessBlue]{AKF}
\definechangesauthor[color=Orange]{AN}
\definechangesauthor[color=brown]{AK}

\usepackage{mathtools} 

\newcommand{\coop}{\mathcal{C}}

\newcommand{\mec}{\mathrm{m}}

\newcommand{\kappamain}{\kappa}
\newcommand{\kappadba}{\kappa_{\mathrm{DBA}}}
\newcommand{\kappaextmain}{\kappa_{\mathrm{ex}}}
\newcommand{\kappaintmain}{\kappa_{\mathrm{0}}}

\newcommand{\omegaaux}{\omega_{\mathrm{aux}}}
\newcommand{\omegamain}{\omega_{\mathrm{c}}}

\newcommand{\omegam}{\Omega_{\mathrm{m}}}
\newcommand{\gammam}{\Gamma_{\mathrm{m}}}
\newcommand{\gammaeff}{\Gamma_{\mathrm{eff}}}

\newcommand{\inj}{\mathrm{inj}}
\newcommand{\mas}{\mathrm{mas}}
\newcommand{\pinj}{P_\inj}
\newcommand{\pmas}{P_\mas}
\newcommand{\ppump}{P_\mathrm{pump}}

\newcommand{\ominj}{\omega_\inj}
\newcommand{\ommas}{\omega_\mas}
\newcommand{\deltaominj}{\Delta \omega_\inj}

\begin{document}

\title{A maser based on dynamical backaction on microwave light}

\author{L.~D.~T\'{o}th}
\thanks{These authors contributed equally to this work}
\author{N.~R.~Bernier}
\thanks{These authors contributed equally to this work}
\affiliation{Institute of Physics, {\'E}cole Polytechnique F{\'e}d{\'e}rale de Lausanne,
	Lausanne 1015, Switzerland}
\author{A.~K.~Feofanov}
\email{alexey.feofanov@epfl.ch}
\affiliation{Institute of Physics, {\'E}cole Polytechnique F{\'e}d{\'e}rale de Lausanne,
	Lausanne 1015, Switzerland}
\author{T.~J.~Kippenberg}
\email{tobias.kippenberg@epfl.ch}
\affiliation{Institute of Physics, {\'E}cole Polytechnique F{\'e}d{\'e}rale de Lausanne,
	Lausanne 1015, Switzerland}
\date{\today}
\begin{abstract}
  The work of Braginsky introduced radiation pressure dynamical backaction, in which a mechanical oscillator that is parametrically coupled to an electromagnetic mode can experience a change in its rigidity and it's damping rate.
  The finite cavity electromagnetic decay rate can lead to either amplification or cooling of the mechanical oscillator,
  and lead in particular to a parametric oscillatory instability,
  associated with regenerative oscillations of the mechanical oscillator,
  an effect limiting the circulating power in laser gravitational wave interferometers.
 These effects implicitly rely on an electromagnetic cavity whose dissipation rate vastly exceeds that of the mechanical oscillator,
 a condition naturally satisfied in most optomechanical systems.
  Here we consider the opposite limit,
  where the mechanical dissipation is engineered to dominate over the electromagnetic one, essentially reversing role of electromagnetic and mechanical degree of freedom.
  As a result, the electromagnetic field is now subject to dynamical backaction:
  the mechanical oscillator provides a feedback mechanism which modifies
  the damping rate of the electromagnetic cavity.
  We describe this phenomenon in the spirit of Braginsky's original description, invoking finite cavity delay and highlighting the role of dissipation.
 Building on previous experimental work,
 we demonstrate this dynamical backaction on light in a superconducting microwave optomechanical circuit.
 In particular, we drive the system above the parametric instability threshold of the microwave mode, leading to maser action and
 demonstrate injection locking of the maser, which stabilizes its frequency and reduces its noise.
\end{abstract}
\maketitle
More than 50 years ago, the seminal work of Braginsky%
~\cite{braginski_ponderomotive_1967,braginsky_investigation_1970} 
introduced the notion of radiation pressure dynamical backaction,
in the context of a mechanical oscillator coupled parametrically to an electromagnetic mode.
This arrangement enables to measure mechanical motion with high precision as required in particular for gravitational wave detectors,
but
radiation pressure can constitute a limitation.
As the cavity field adjusts to the oscillator motion,
the radiation pressure force it generates acts as a feedback force
which can acquire an out-of-phase component
due to the finite cavity delay and modify the mechanical damping rate.
This dynamical backaction poses
a limitation to the circulating power in Fabry-Perot interferometers,
due the \emph{parametric oscillatory instability},
in which amplification compensates the intrinsic mechanical losses,
leading to regenerative oscillations of the mechanical end mirror.
Radiation pressure parametric instability, as proposed by Braginsky%
~\cite{braginsky_parametric_2001}
 was first observed in toroid micro-resonators in 2005~\cite{kippenberg_analysis_2005},
 and gives rise to a rich nonlinear dynamics%
~\cite{marquardt_dynamical_2006}.
Soon thereafter, dynamical backaction cooling, an effect Braginsky predicted
~\cite{braginsky_low_2002}
to occur for red-detuned laser excitation, was demonstrated%
~\cite{gigan_self-cooling_2006,arcizet2006,schliesser_radiation_2006}.
Although the parametric instability was first analyzed for a single electromagnetic mode coupled to a mechanical oscillator, the effect can also occur for multi-mode systems in which modes are spaced by the mechanical frequencies%
~\cite{kells_considerations_2002},
a scenario in which the parametric oscillator stability
in advanced LIGO at the Livingston observatory%
~\cite{evans_observation_2015} was observed.
Although undesirable in the context of LIGO, the ability to amplify and cool mechanical motion using dynamical backaction is at the heart of the advances in cavity opto- and electromechanics over the past decade~\cite{cavity_optomechanics_RMP} that have enabled mechanical systems to be controlled at the quantum level.
Dynamical backaction control over mechanical oscillators has enabled to cool micro- and nanomechanical oscillators to unprecedentedly low entropy states~\cite{verhagen_quantum-coherent_2012,teufel_sideband_2011,chan_laser_2011},
and thereby opened a path to study optomechanical quantum effects ranging from optomechanical squeezing, mechanical squeezed states, sideband asymmetry, to entanglement of mechanical motion with microwaves%
~\cite{safavi-naeini_squeezed_2013,purdy_strong_2013,sudhir_appearance_2017,nielsen_multimode_2017,wollman_quantum_2015,palomaki_entangling_2013,pirkkalainen_squeezing_2015,lecocq_quantum_2015,riedinger_non-classical_2016}.

\begin{figure*}
  \includegraphics[width=0.7\textwidth]
  {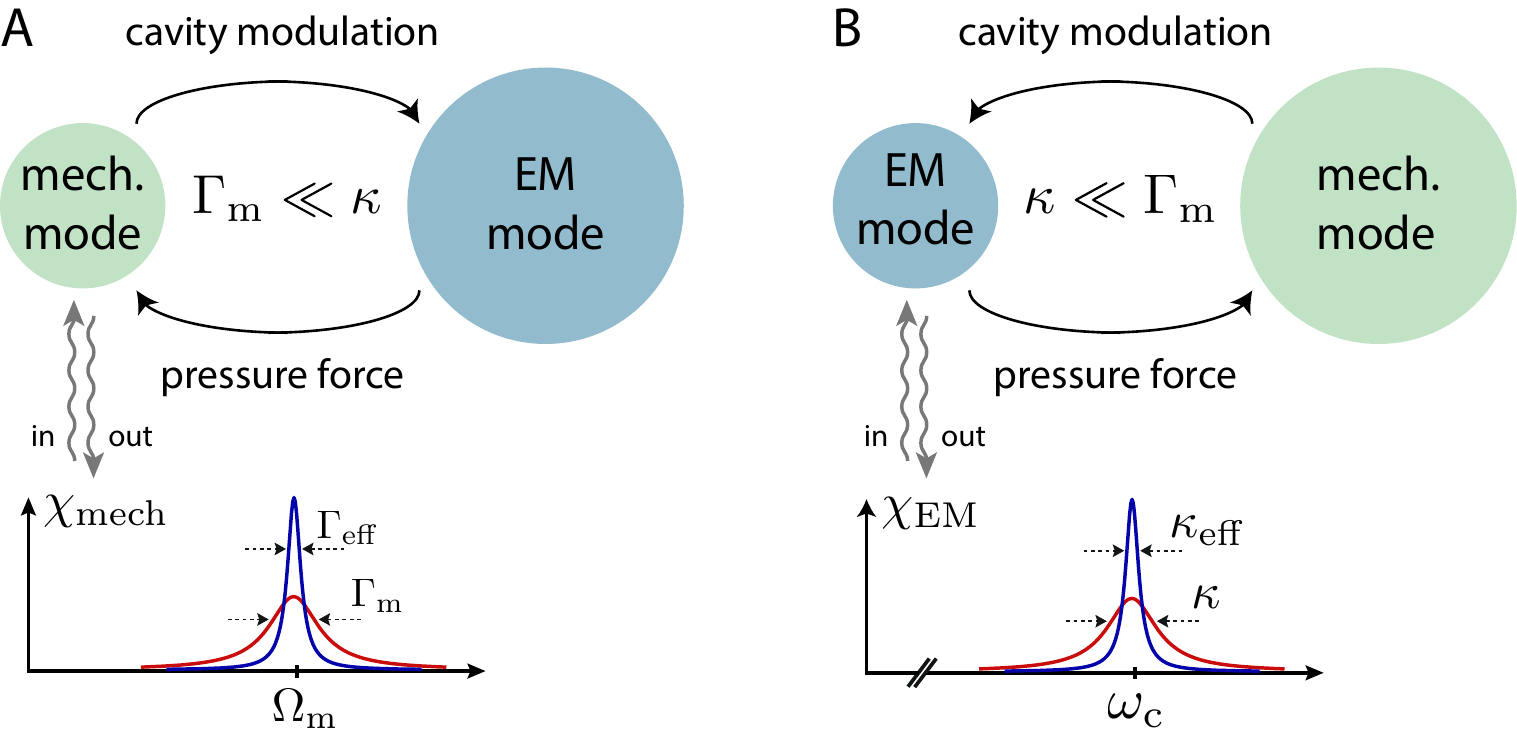}
  \caption{
  \textbf{The role of dissipation in dynamical backaction.}
  \textbf{A}.~An electromagnetic mode (with resonant frequency $\omegamain$ and energy dissipation rate $\kappamain$) is coupled to a mechanical oscillator (with resonant frequency $\omegam$ and energy dissipation rate $\gammam$) through the optomechanical interaction, in which the former exerts radiation pressure force, while the latter modulates the resonant frequency of the cavity.
  In standard optomechanical systems, the dissipation rates satisfy the hierarchy $\kappamain \gg \gammam$ and this interaction can be viewed as a simple feedback mechanism acting on the mechanical oscillator.
  This modifies the oscillator's damping rate,
  in a process coined as ``dynamical backaction'' by Braginsky.
  \textbf{B}.~In the scenario where the optical mode is coupled to a mechanical oscillator whose dissipation rate dominates over that of the optical mode, the role of the modes is reversed.
  Here, the mechanical mode provides the feedback mechanism (dynamical backaction) for the optical mode, therefore modifying the light mode's resonant frequency and damping rate.
  \label{fig:scheme}
   }
\end{figure*}

It is interesting to highlight the role of dissipation in dynamical backaction.
Indeed dissipation
determines the resulting modification of the mechanical and optical susceptibility
due to the optomechanical interaction.
In almost all optomechanical systems ranging from gravitational wave observatories to nano-optomechanical systems,
the electromagnetic dissipation dominates over the mechanical one,
leading to the above mentioned optomechanical phenomena.
In contrast, if the mechanical oscillator is more dissipative than the electromagnetic mode,
the roles are reversed~\cite{nunnenkamp2014}.
In this situation,
(but still obeying the condition that the mechanical dissipation occurs on a timescale that is long compared to the mechanical oscillator period, or equivalently stated the mechanical quality factor is still exceeding unity),
the dynamical backaction that occurs for detuned laser excitation causes
a feedback force that is applied to the \emph{electromagnetic} mode,
resulting in amplification or damping.
This \emph{electromagnetic dynamical backaction} leads to
a parametric oscillatory instability
that now
corresponds to the action of a \emph{maser}.
(i.e. the stimulated emission of microwaves).
Here we describe an experiment in which we realize
such a maser based on
the dynamical backaction amplification of microwave light.
We design
a microwave optomechanical circuit~\cite{teufel_circuit_2011}
with two microwave cavities coupled to the same mechanical oscillator, formed by a vacuum gap capacitor~\cite{cicak_low-loss_2010}.
The auxiliary (low Q) microwave mode is used to sideband-cool the mechanical mode
until its dissipation rate dominates over the main (high Q) microwave mode.
With a strong blue-detuned tone, the main mode is coupled to the mechanical mode,
resulting in microwave gain from dynamical backaction and masing.
We proceed to perform injection locking of this maser.

\begin{figure}
  \includegraphics[width=0.4\textwidth]
  {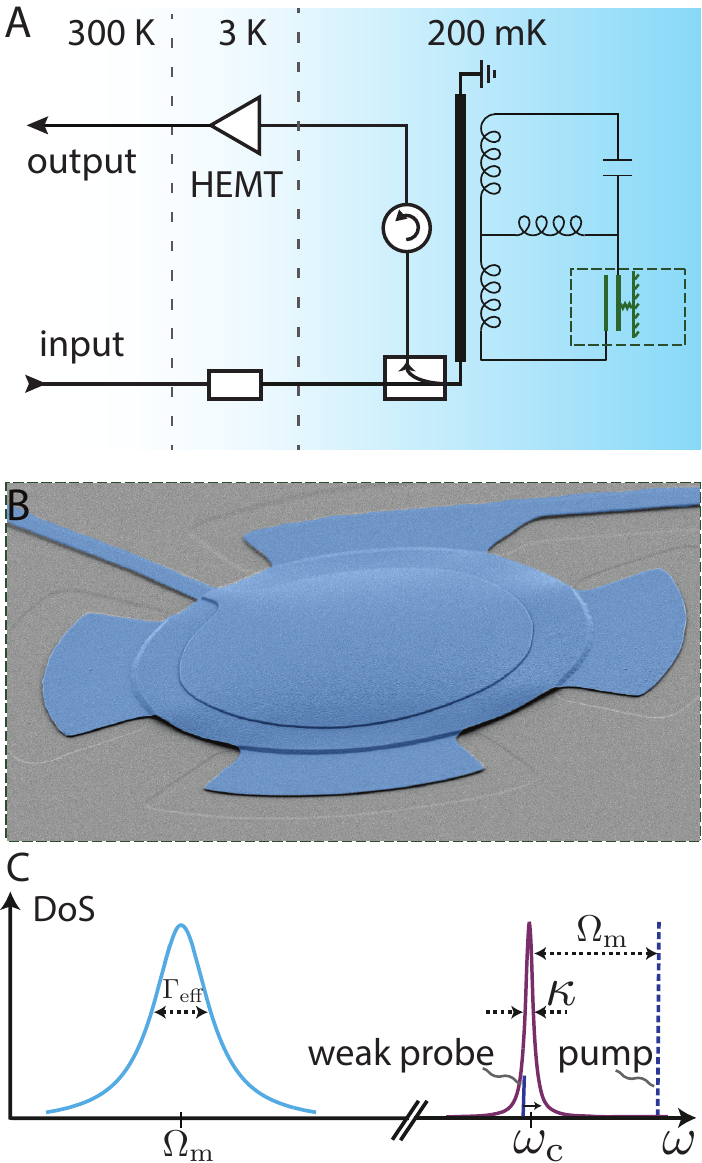}
  \caption{
  \textbf{Experimental setup and measurement scheme.}
  In order to observe dynamical backaction on a light mode,
  a superconducting microwave circuit with a mechanically compliant vacuum gap capacitor~\cite{cicak_low-loss_2010} is fabricated
  and measured in a dilution refrigerator.
  \textbf{A}.~To pump and probe the superconducting microwave circuit,
 microwave sources and a network analyser are combined at room temperature and connected to the
  circuit through various filters and attenuators at different temperature stages (see Ref.~\cite{toth_dissipative_2016} for details).
  The reflected signal at the output is amplified and measured
  either by an electromagnetic spectrum analyser 
  or a vector network analyser.
  \textbf{B}.~Electron micrograph of the mechanically compliant vacuum-gap capacitor.
  \textbf{C}.~The mechanical oscillator is electromagnetically damped using an auxiliary mode (not shown) such that $\gammaeff \gg \kappamain$.
  Once this is achieved, a pump is placed on the upper motional sideband of the electromagnetic mode.
  In the injection locking experiment,
  a weak tone is applied near the resonance frequency $\omegamain$.
  \label{fig:exp}
   }
\end{figure}

\section*{Mathematical description of Dynamical Backaction}
We start by reformulating
Braginsky's original derivation of dynamical backaction%
~\cite{braginski_ponderomotive_1967,braginsky_parametric_2001}
for the case of a blue-detuned pump,
in a way that showcases how the process is reversed in the case of an opposite dissipation hierarchy.
Consider an electromagnetic mode with frequency $\omegamain$ and
energy dissipation rate $\kappamain$ coupled, via the standard optomechanical coupling,
to a mechanical mode with
resonant frequency $\omegam$ and energy dissipation rate $\gammam$.
While Braginsky considered the limit
$\omegam \leq \kappamain$
to derive the delayed feedback force experienced
 by the mechanical oscillator
due to the electromagnetic field,
we consider here the sideband-resolved regime%
~\cite{marquardt2007,wilson-rae2007,schliesser_resolved-sideband_2008}
$\omegam \gg \kappamain$,
relevant for our experiment.
We consider the equations of motion parametrically coupling
the mechanical oscillator to the electromagnetic cavity.
These (for simplicity classical and linearized) equations can be written in a rotating frame
for the phasors of the two modes;
in the case of a pump driving the system on
the upper motional sideband at a detuning
\newcommand{\forcea}{F_a}
\newcommand{\forceb}{F_b}
$\Delta = \omegam$, they are given by the coupled-modes equations
\begin{align}
  \dot a^* (t)
  &=
  -\frac{\kappamain}{2} a^* (t)
  + i g b(t)
  =
  -\frac{\kappamain}{2} a^* (t)
  - i \forcea(t)
  \label{eq:motion1}
  \\
  \dot b (t)
  &=
  -\dfrac{\gammam}{2} b(t)
  - i g a^*(t)
  =
  -\dfrac{\gammam}{2} b(t)
  + i \forceb(t),
  \label{eq:motion2}
\end{align}
where
$b (t) = (\sqrt{m\omegam} x(t) + i \sqrt{1/m\omegam} p(t))/\sqrt{2}$
describes the state of the mechanical mode of mass $m$,
$a(t)$ is the phasor for the electromagnetic mode
and $g = g_0 |\alpha_0|^2$ is the vacuum optomechanical coupling rate $g_0$
enhanced by the field $\alpha_0$ of the blue-detuned pump.
In the rotating frame, the variables describe the slowly changing
amplitude and phase of the rapidly oscillating field and oscillator,
at $\omegamain$ and $\omegam$.
Each of the two harmonic oscillators is subject to a ``force'' 
(denoted by $\forcea(t)$ and $\forceb(t)$) 
proportional to the state of the other harmonic oscillator, establishing a feedback mechanism:
cavity intensity fluctuations create a radiation pressure force acting
on the mechanical oscillator, while
mechanical displacement modulates the cavity resonance frequency.
The symmetry of the relationship is broken by the different scales
of the dissipation rates.

Braginsky originally considered the case
where electromagnetic dissipation dominates
($\kappamain \gg \gammam$).
This is natural in most systems as the quality factors are commensurate for the electromagnetic and mechanical modes,
while there is a large (many orders of magnitude) separation of scales in their respective frequencies.
In this limit, the electromagnetic field envelope
almost instantly adapts to the mechanical displacement
($\dot a^*(t) \approx 0$)
and becomes proportional to it such that
$a^*(t) = i(2g/\kappamain) b(t)$.
The field then exerts a force on the mechanical oscillator
proportional to the state of the latter, given by
$\forceb(t) = -i(2g^2 / \kappamain) b(t)$.
Therefore, the interaction can be viewed as a simple feedback loop.
The factor $-i$ represents a delay of a quarter period
for the feedback force acting on the mechanical oscillator.
This delay means that the force,
acting in quadrature,
increases the amplitude of the phasor,
equivalent to a decrease in mechanical damping or gain.
If the pump detuning $\Delta$ does not fall exactly on the sideband,
the delay is not exactly $i$ and the force has an in-phase component,
modifying the frequency of the mechanical oscillator
(this effective change of the mechanical spring constant
is called the \emph{optical spring} effect).
The amplification process can be understood as a positive feedback that measures
the state of the mechanical oscillator and returns it with a delay
as a force,
i.e. a dynamical backaction
(see \cref{fig:scheme} A).
For sufficiently high coupling strength $g$,
this leads to a parametric oscillatory instability, that causes regenerative oscillations of the mechanical oscillator, and thus
limits the maximal circulating power for a gravitational wave detector%
~\cite{braginsky_parametric_2001}.

In this letter, we study the converse process,
where the mechanical dissipation rate dominates
($\gammam \gg \kappamain$).
Now the envelope of mechanical oscillations nearly instantly adjusts
to the state of the electromagnetic field
($\dot b(t) \approx 0$)
and is proportional to it
such that
$b(t) = -i(2g / \gammam) a^*(t)$.
The field is then subject to
an in-quadrature ``force''
proportional to its own state
$\forcea(t)
=
i( |\kappadba| / 2)
a^{*}(t)
$
where $\kappadba = -4 g^2 / \gammam = -\kappa \coop$,
introducing the multiphoton cooperativity
$\coop = 4 g^2/(\kappamain \gammam)$.
Similarly to above, the force has
a delay $+i$ of a quarter period and
increases the amplitude
of electromagnetic oscillations, compensating for damping
by an amount given by $\kappadba$,
such that the effective energy decay rate of the cavity is
$\kappamain + \kappadba$.
A change in detuning $\Delta$ would again
slightly modify this delay and
create components
of the in-phase force component, changing
the speed of oscillations, and displacing the resonance frequency
of the cavity (thus creating a \emph{mechanical spring} effect).
This is equivalent to a feedback loop for
the electromagnetic mode
(see \cref{fig:scheme} B),
and implies that the mechanical oscillator is responsible
for \emph{dynamical backaction on light}.
As above, the positive feedback can lead to a parametric instability.
For $\kappadba=-\kappamain$,
the anti-damping caused by this feedback exactly compensates the losses (both intrinsic and external) of the electromagnetic mode, and
the cavity develops self-sustained oscillations
i.e. acts as a maser.
The intrinsic optomechanical nonlinearity sets the
maximum amplitudes of the oscillations%
~\cite{nunnenkamp2014}
and the dynamics is no longer captured by the linearized 
\cref{eq:motion1,eq:motion2}.
Experimentally, this dynamical backaction on the microwave mode 
can be observed by measuring the emission spectrum of the electromagnetic mode.
Below the masing threshold, one expects amplified noise at the output of the device 
in a bandwidth which is commensurate with the effective energy decay rate
$\kappamain + \kappadba$.
At the threshold, this bandwidth collapses to zero and a strong, spectrally pure signal emerges from the cavity.

\section*{Experimental realization}
We experimentally explore dynamical backaction on an electromagnetic (EM) mode from a mechanical oscillator
and the resulting maser action in a microwave optomechanical circuit.
The circuit, made of thin-film aluminium on a sapphire substrate, supports two EM modes
in the microwave regime (we denote them as primary and auxiliary modes
with resonance frequencies $\omegamain = 2\pi \cdot 4.08$~GHz and $\omegaaux = 2\pi \cdot 5.19$~GHz, respectively).
One of the elements in the circuit is a parallel-plate vacuum-gap capacitor with a suspended top electrode,
forming the mechanical oscillator (\cref{fig:exp} B).
The resonance frequency of the fundamental flexural mode is
$\Omega_\mec=2\pi \cdot 6.5 \; \mathrm{MHz}$, whose motion is
coupled to both EM modes with
a vacuum electromechanical coupling strength
$g_0 \approx 2 \pi \cdot 60\;\rm{Hz}$.
The device is cooled to 200 mK in a dilution refrigerator (see \cref{fig:exp} A) and
can be probed in reflection using a vector network analyzer or its spectrum can be measured
at the output port using a spectrum analyzer.
The input line is filtered and attenuated at various temperature stages to eliminate extraneous
Johnson- and phase-noise and the output line is amplified using a commercial HEMT amplifier at 3 K.
Using a strong pump on the lower motional sideband of the auxiliary EM mode,
the mechanical mode is damped to $\gammaeff = 2\pi \cdot 440$ kHz,
or $\sim 2.5$ times the energy dissipation rate of the primary EM mode.

\begin{figure}[b]
  \includegraphics
  {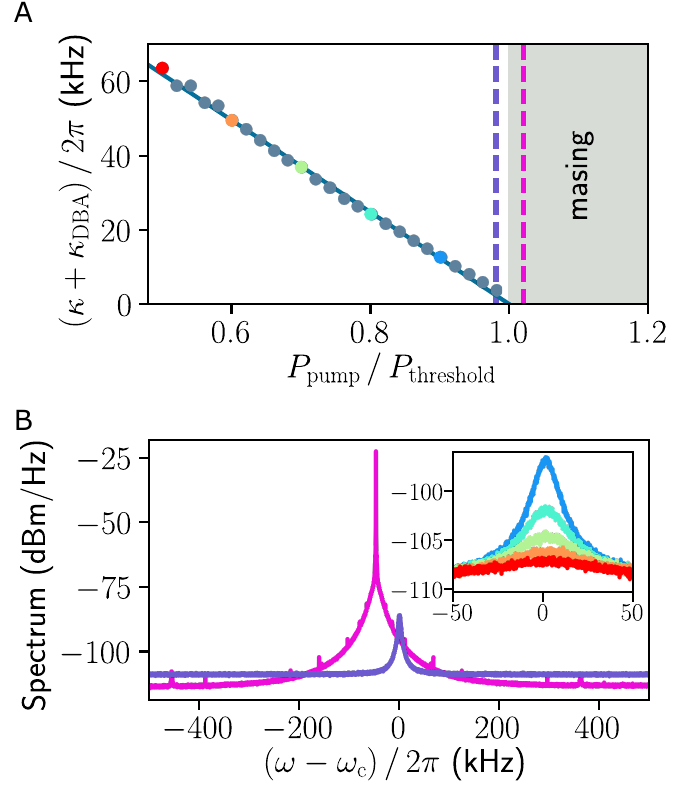}
  \caption{
  \textbf{
  Dynamical backaction amplification and the masing threshold.
  }
  \textbf{A}.~The electromagnetic mode is coupled to the dissipative mechanical mode by a pump placed on the upper motional sideband.
  As the pump power $\ppump$ is increased,
  the apparent width of emission
  $\kappamain + \kappadba$
  of the electromagnetic mode
  (corresponding to its effective energy decay rate) decreases
  as the dynamical backaction antidamping
  $|\kappadba|$
  increases linearly with power.
  When $\kappadba$ compensates for the intrinsic dissipation rate $\kappamain$,
  this positive feedback leads to a parametric instability and masing.
   \textbf{B}.~Output spectrum of the electromagnetic mode
   just below  and above threshold, corresponding to the dashed lines in panel A.
   The inset shows the amplification and narrowing of emission below threshold
   with pump power, with each trace corresponding to the matching colour dot in panel A.
  \label{fig:masing}
   }
\end{figure}
\begin{figure*}[t]
  \includegraphics
  {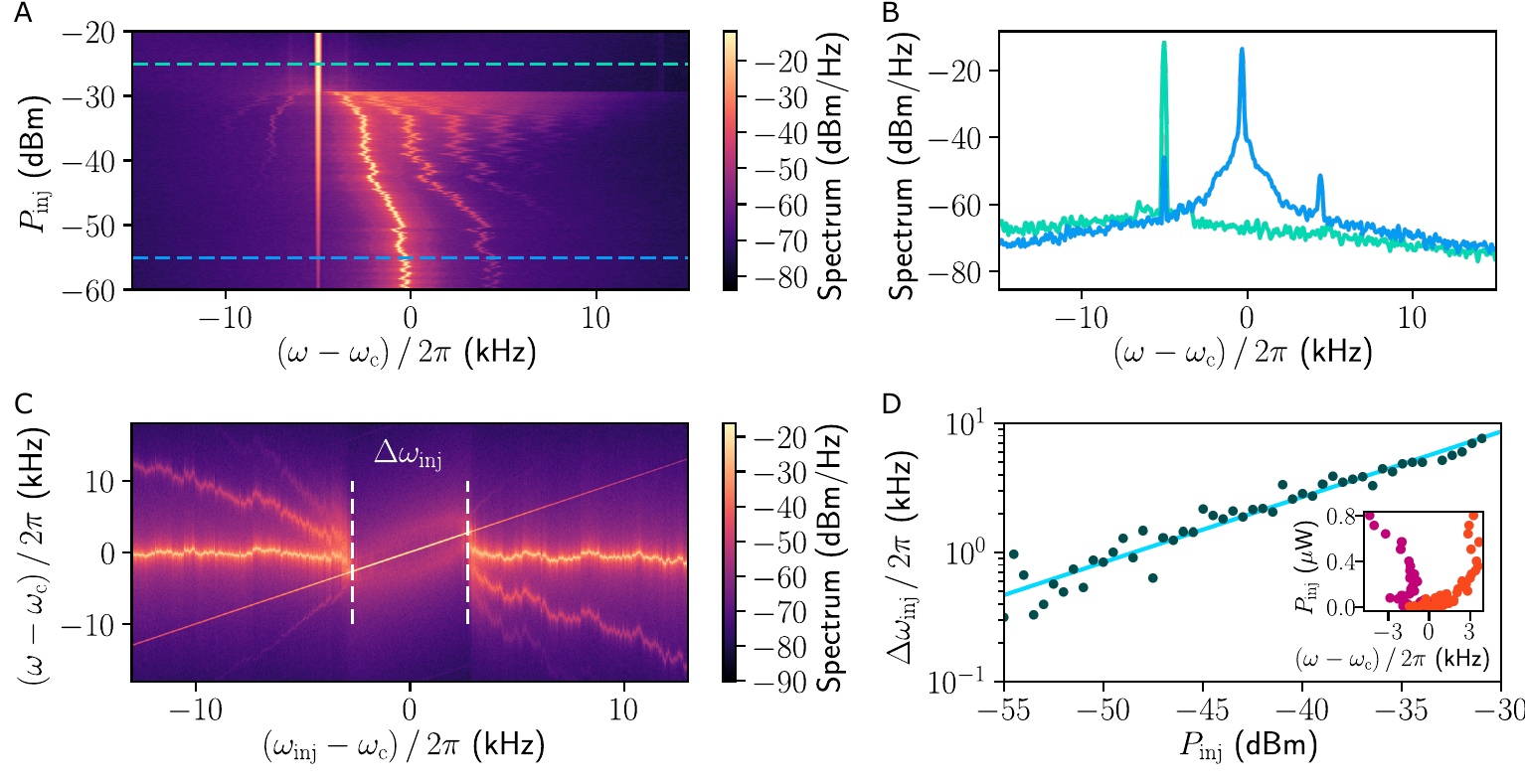}
  \caption{
  \textbf{Injection locking of a maser based on dynamical backaction}
  \textbf{A}.~Output spectrum of the maser, as the power $\pinj$
  of an injected tone  detuned 5 kHz to the red is increased. The power is measured at the output of the corresponding microwave source.
  Above a threshold power of about -30 dBm, the maser locks to the tone, considerably suppressing the noise and eliminating the frequency jitter present in the free-running case.
  The intrinsic nonlinearity results in distortion sidebands 
  from the two tones beating below threshold.
  \textbf{B}.~Spectra corresponding to cuts of A along the dashed lines,
  below and above the locking threshold.
  \textbf{C}.~Output spectrum of the maser as
  the frequency $\ominj$ of a weak tone
  of constant power $\pinj = -40$ dBm
  is swept across the free-running maser frequency $\ommas$.
  The locking range $\deltaominj$, wherein
  the two oscillations are frequency locked, is highlighted.
  \textbf{D}.~The locking range $\deltaominj$
  as a function of the injected tone power $\pinj$.
  A fit on the logarithmic scale gives a slope of 0.51, 
  confirming the expected scaling
  $\deltaominj \propto \sqrt{\pinj}$.
  The inset depicts the actual limiting points of the locking range as a function
  of tone power $\pinj$, illustrating the Arnold tongue of the system.
  \label{fig:lock}
   }
\end{figure*}
Then, we couple the mechanical motion to the primary EM mode using a pump placed
on the upper motional (blue) sideband (see \cref{fig:exp} C).
This coupling introduces dynamical backaction on the microwave mode,
changing its susceptibility (i.e. resonance frequency and damping rate) as we increase the pump power.
A prominent manifestation of this change, in the case of a pump on the upper motional sideband,
is the narrowing of the linewidth of the electromagnetic resonance, corresponding to a decrease in the energy decay rate (anti-damping).
The apparent decay rate of the cavity becomes
$\kappa + \kappadba$, where
$\kappadba = -\coop \kappamain$
is linearly proportional to the power of the tone at the blue sideband.
We monitor the output spectrum of the cavity
for different pump powers, as shown in \cref{fig:masing} A
and observe that the width of the resonance
decreases in a linear fashion, as expected.
Above a certain pump power, the feedback mechanism introduced by dynamical backaction compensates the intrinsic loss $\kappaintmain$
of the microwave mode and the system acts as an amplifier with net gain
(the inset of \cref{fig:masing} B shows amplified noise emerging from the cavity for different pump powers).
At sufficiently high pump power, for unit cooperativity $\coop=1$,
dynamical backaction compensates
the total energy decay rate of the EM mode
($|\kappadba| = \kappamain = \kappaintmain + \kappaextmain$,
where $\kappaextmain$ is the external coupling rate),
and the system undergoes a transition into the self-sustained oscillatory regime.
The microwave cavity, now acting as a maser,
emits a single, pure tone at its resonance frequency (\cref{fig:masing} B),
orders of magnitude ($\sim 55$ dB) stronger than just below the threshold.

\section*{Injection locking}
We now demonstrate the locking of our maser
with a weak injected tone.
Injection locking is a synchronization phenomenon
of lasers and masers~\cite{siegman_lasers_1986},
and has been demonstrated in many systems,
including recently in
a trapped-ion phonon laser~\cite{knunz_injection_2010},
a quantum cascade laser~\cite{st-jean_injection_2014},
a quantum-dots maser~\cite{liu_injection_2015}
as well as an AC Josephson junction maser~\cite{cassidy_demonstration_2017}.
A weak tone of frequency $\ominj$
close to the maser emission frequency $\ommas$
will compete for gain with it in a way
that effectively couples the two oscillations and permits synchronization.
The phenomenon is generally described by the
Adler equation~\cite{adler_study_1946}
\begin{equation}
  \frac{d \phi}{d t}
  + \left( \ominj - \ommas \right)
  =
  - \frac{1}{2} \deltaominj \sin \left( \phi \right)
  \label{eq:adler}
\end{equation}
which models the dynamics of the relative phase $\phi$
of the two oscillations.
If the injected tone falls within a locking range of width
$
\deltaominj
=
2 \kappaextmain \sqrt{\alpha \pinj / \pmas}
$ centered around the masing frequency $\ommas$,
the two tones lock and the phase difference $\phi$ becomes constant.
This range depends on the ratio between the injected tone power $\pinj$,
attenuated by factor $\alpha$ at the input of the cavity,
and the maser emission power $\pmas$.
Outside this range, the Adler equation predicts 
that the maser frequency is pulled towards the injected tone and 
that  distortion sidebands appear due to the two tones beating
and the intrinsic nonlinearity%
~\cite{adler_study_1946}.
We first study this phenomenon by placing the injected tone
5 kHz away from the maser and monitoring the output spectrum
while the injected power $\pinj$ is varied
(\cref{fig:lock}~A).
The maser emission is pulled towards the injected tone
and finally locks at an injected power threshold corresponding here
to about -30 dBm.
As the two tones become comparable in strength,
distortion sidebands from the beating increase in amplitude.
In the locked region, the noise surrounding the peak  is considerably suppressed
compared to the free-running case and the frequency jitter (originating from frequency instability of the cavity and the mechanical mode) is eliminated
(\cref{fig:lock}~B).
We proceed to measure the locking range $\deltaominj$,
by fixing the injected power $\pinj$
and sweeping its frequency $\ominj$
across the maser frequency
(\cref{fig:lock}~C).
When the frequency difference is below $\deltaominj$,
the two tones lock.
The noise around the peak is suppressed and
the frequency jitter of the maser ceases.
Repeating the measurement at different injected powers $\pinj$,
the locking range
is shown to obey the expected scaling law
$\deltaominj \propto \sqrt{\pinj}$
(\cref{fig:lock}~D).
Finally, as an inset,
the limit points of locking are shown as a function of power,
drawing the so-called Arnold tongue%
~\cite{pikovsky_synchronization_2003}.
The asymmetric shape is due to drift of the masing frequency $\ommas$
during the measurement, which do not affect
the locking range $\deltaominj$.
\section*{Summary}
In conclusion, we have described how dynamical backaction relies on the dissipation hierarchy between an electromagnetic (e.g. optical or microwave) mode and a mechanical mode. By modifying and reversing this hierarchy of dissipation dynamical backaction occurs on the \emph{electromagnetic mode}, leading to amplification of microwaves, and eventually an instability, i.e. maser action.
We realize and exploit this novel backaction mechanism in a superconducting circuit electromechanical system.
When the system is pumped on the upper motional sideband of the cavity mode, backaction provides a positive feedback loop,
leading to amplification and, when this feedback compensates the total energy decay rate of the cavity,
to parametric instability.
This results in maser action: a strong, spectrally pure tone emerges at the resonance frequency of the microwave cavity.
We have measured injection locking of the maser, reducing its noise and stabilizing its frequency and have shown that the injection locking range as a function of input power follows
Adler's theory.

\begin{acknowledgments}
The authors declare no competing financial interests. This work was supported by the SNF,
the NCCR Quantum Science and Technology (QSIT),
 the European Union Seventh Framework Program
through iQUOEMS (grant no.~323924) and Marie Curie ITN cQOM (grant no.~290161).
TJK acknowledges financial support from an ERC AdG (QuREM).
All samples were fabricated in the Center of Micro\-Nano\-Technology (CMi) at EPFL.
\end{acknowledgments}

\bibliography{bibliography}
\end{document}